# Using Amazon Alexa APIs as a Source of Digital Evidence

Clemens Krueger
*School of Computing*
*Edinburgh Napier University*
Edinburgh, Scotland
clemens@f-krueger.net

Sean McKeown
*School of Computing*
*Edinburgh Napier University*
Edinburgh, Scotland
s.mckeown@napier.ac.uk

*Abstract*—With the release of Amazon Alexa and the first Amazon Echo device, the company revolutionised the smart home. It allowed their users to communicate with, and control, their smart home ecosystem purely using voice commands. However, this also means that Amazon processes and stores a large amount of personal data about their users, as these devices are always present and always listening in peoples' private homes. That makes this data a valuable source of evidence for investigators performing digital forensics. The Alexa Voice Service uses a series of APIs for communication between clients and the Amazon cloud. These APIs return a wide range of data related to the functionality of the device used.

The first goal of this research was to clarify exactly what kind of information about the user is stored and accessible through these APIs. To do this, a combination of literature review and exploratory analysis was used to establish a list of all relevant APIs. Then, possible artefacts and conclusions to be drawn from their responses were identified and presented. Lastly, the perspective of the users was taken, and options for improving their privacy were reviewed. Specifically, the history of interaction between the user and Alexa is available through multiple APIs, and there are several options to delete it. It was determined that these options have different behaviours and that most of them do not remove all data related to user interaction.

## I. Introduction

Recently, the usage of devices connected to the Internet of Things (IoT) has increased significantly. A lot of these devices can be attributed to the concept of the "smart home", which is intended to provide convenience in day-to-day activities. This leads to the fact that they tend to record and sometimes store a lot of personal data belonging to their users. This data often not only describes the user directly, but also their daily schedules and their habits. In the case of wearable smart devices, critical medical information may even be processed, stored, and transmitted. All of this is highly sensitive information, which should be kept private and secured appropriately.

Within the smart home, there is one type of device in particular that stands out, which is the virtual personal assistant, such as Google Home or Amazon Echo with the Alexa assistant service. They are meant to be the user's interface to connect to their smart home and can pick up voice commands and interpret them accordingly to control other devices and services, either by using voice commands or scheduled actions.

Therefore, the device turns into a hub that is connected to every smart device in the house, which in turn also means that it gets access to a variety of personal data about its users. Examples could be identifying sleep habits by checking when light bulbs are turned on, deducing a person's health from examining heart rates from a smartwatch, or more personal traits extracted from recordings of conversations.

To be able to provide basic functionality, virtual personal assistants necessarily need to record audio and process it. In the case of Amazon Alexa for example, the device listens for someone to say the word "Alexa", which is its trigger to start recording and processing. Therefore, Alexa needs to constantly record audio to be able to pick up the command. Because interpreting audio is a complex challenge, the processing is usually not performed on the local device. Rather, the recordings are sent to Amazon's servers, where the bulk of processing takes place.

Unfortunately, since these systems are all closed source, it is difficult to determine what processing is applied to data gathered by Alexa. While it is clear that at least some of the data is sent to Amazon, it is unknown how the company then handles this information. Bloomberg recently discovered [1] that Amazon employees had access to voice recordings of customers, and actively listened to them to improve Alexa's voice recognition system. According to the source, employees "transcribed and annotated" the recordings, and sometimes even heard conversations. In an extreme case, two employees reportedly even listened to a case of sexual assault. Other virtual assistant devices face similar problems. The Guardian [2] reports that certain contractors of Google have access to voice recordings. This was admitted by Google following a recent incident where some of these recordings were leaked. This is especially critical considering that the Amazon Echo device is usually placed directly in peoples' homes, and can therefore record a lot of very private details and conversations.

According to Gartner [3], the adoption of such devices will increase drastically in the next few years. While in 2016 the market value of virtual personal assistant devices was at $720 million, it is projected to rise to $3.52 billion by 2021, almost five times as much. This clearly illustrates how these issues will become very relevant in the future.

Due to the amount of data processed and stored by them,

these devices are becoming more and more relevant for criminal investigations, since they can potentially contain important evidence in cases such as domestic violence. Amazon has already been ordered to hand out data gathered by Echo devices as part of the evidence in criminal trials. In the US, a judge ordered Amazon to hand over data from an Echo device that was used in a house where a double homicide was committed in January 2017 [4]. Specifically, existing voice recordings as well as information on cellular devices that were paired with the Echo during this time. The same thing occurred in 2015 when another Echo device was seized during the investigation of a suspected murder in Arkansas [5]. In this case, Amazon only agreed to hand over data after the suspect had agreed to it. These investigations make use of digital forensics techniques in order to preserve potential pieces of evidence from these devices, and to identify the parts that allow conclusions about a certain situation.

## II. AMAZON ECHO & AMAZON ALEXA

The Amazon Echo is a family of intelligent virtual assistant devices that come in different form factors and with different featuresAt the core, however, they all contain a microphone and a speaker and interact with the Amazon Alexa voice service which allows the user to control it using voice commands.

As such, they combine technologies related to the Internet of Things as well as cloud computing. As mentioned before, these devices have recently started to raise interest as sources of digital evidence for criminal investigations. However, because the technology is relatively new, research on how best to extract forensics evidence from such devices is still in its early stages. This section will provide an overview of the devices' functionality, as well as a summary of the current state of research in the area of Amazon Echo forensics.

The Echo product family consists of several products, among which are the classic Echo, the smaller Echo Dot and the Echo Show, which also contains a small display to show additional information. The Echo is capable of performing several tasks that are meant to support users during their daily lives. It is possible to ask the device questions to search on the Internet, it can convert scientific units and currencies, present the current time and weather, and more. Additionally, one can have it set timers and reminders and create to-do or shopping lists. Another big aspect is that the Echo can also control third-party hardware and software that is connected to it. It can control light bulbs or central heating system, as well as play music from several streaming services.

Lastly, the user can also define so-called skills. These skills can be invoked by specific voice commands and can trigger arbitrary actions by the Echo. For example, a certain skill that is invoked by saying "Alexa, set the mood" could prompt the Echo to dim the lights in a room and start playing a certain music playlist.

### A. Network structure

The setup of an Echo device is done through a so-called "companion app", which can either be a specific application on Android or iOS devices, or the Amazon Alexa Web page at https://alexa.amazon.com. During setup, the Echo connects to a local wireless network. After that, it is in constant contact with Amazon's cloud environment. In fact, if the Echo is not connected to the Internet, it does not work, because all the voice commands are processed in the cloud, and not locally on the device.

Voice commands to the device are triggered using a specific wake word, such as "Alexa", or "Echo". When this word is said, the device starts recording audio and transmits the recording to the Alexa Voice Service (AVS) for processing, which then returns an appropriate response to the device, where it is returned to the user through the speaker [6]. This means that the device's microphone is always turned on, as it needs to be able to detect the wake word. According to Amazon, however, no recordings are saved without the wake word [7]. Additionally, all data pertaining to the Echo can be managed through the companion apps mentioned before. After the setup is complete, the Echo device and the companion apps do not communicate directly with each other. They are both connected to the Amazon cloud, which acts as a communication proxy.

### B. Forensic Approaches

After identifying devices that might be of interest in an investigation, the investigator must then acquire data from them, which might potentially contain relevant pieces of evidence [8].

Timelines of events that occurred during the situation in question are among the most valuable pieces of evidence, especially if they include information such as timestamps or location details [9]. If timestamps are involved, problems can arise regarding different time zones. It is therefore important to consider that different data sources may yield data with different time zones or formats, which should then be normalised. To achieve this, certain tools can be employed that visualise timelines in a graphical format. This can give the investigator an overview of the relevant events in a way that allows them to see the big picture. These events can come from many different sources, including creating, modifying or deleting a file, log file entries, network activity and more.

There have been several approaches to forensically analyse Amazon Echo that yielded detailed information about how the device communicates with the Amazon cloud and what types of data are processed. Orr and Sanchez [10] performed a study to confirm whether data processed by an Amazon Echo device could be a valuable source of evidence for criminal investigations. In order to achieve this, they had a family use an Echo device in their daily lives for 10 weeks. Afterwards, they extracted information mainly from the Amazon Alexa Web page. The conclusion to their analysis was that the data can indeed contain relevant artefacts for an investigation. Among these artefacts are records of voice interactions between the user and Alexa including timestamps, traffic information with location details as well as to-do and shopping lists. They pointed out that all this evidence can provide valuable information about a suspect. However, it should always be used together with additional sources.

Chung et al. [6] also discussed the topic of forensic approaches to Amazon Alexa. They focused on technical details for methods of analysis. They combined aspects of cloud forensics with client-side forensics by analysing both Alexa API responses and local artefacts. They also proposed a toolkit that facilitates identifying and gathering various forensic artefacts through API calls to the Amazon Alexa service.

Most of the traffic between the Echo and Amazon's servers is encrypted with TLS v1.2. Data exchanges between the two are performed using a predefined API, to which the Echo can send requests that return data in JSON format. There is however no official documentation that contains details about which API calls exist and what data they return. There have been multiple attempts to reverse-engineer this API, which have revealed a number of interesting API calls. Owen Piette published a blog post [11] on his Website explaining how he was able to extract some API calls by looking at which requests his browser sent when he was using the Alexa Web application. Chung et al. [6] then took this further by extensively analysing network communications, browser cache artefacts and data stored by the available companion apps. Through this, they were able to identify almost 20 different APIs which cover all functionality of the Echo device.

During the analysis of the companion apps on Android and iOS, Chung et al. [6] realised that both contained databases with certain cached data sets that were originally obtained by the application through the APIs mentioned above. Also, the authors investigated potential cached values related to the Alexa Web application, where they found several HTTP requests to the API, as well as cached responses.

These investigations allowed Chung et al. [6] to piece together a comprehensive list of APIs that the Echo device uses to transfer and receive data to and from the Amazon Alexa cloud. The researchers defined seven categories, in which they divided the APIs according to the type of data they returned. These categories can be seen in table I. The table contains the name of each category, a short description, as well as examples for the data sets that are part of it. In the original paper, the last category is called *ETC*. For this paper, it was renamed to *Audio Data*, since the only API in this category allows the users to access audio recordings of their voice commands.

Li et al. [12] presented a new process for performing forensic analyses on IoT devices. Like conventional processes, their approach is based on the four stages: identification, preservation, analysis and presentation, all of which are specifically adjusted for the Internet of Things. Following this, they performed a forensic analysis using this process on an Amazon Echo device. Their main sources of evidence are cached artefacts stored by the Amazon Alexa companion app on both Android and iOS, as well as the Alexa Web page. They defined different groups for the data they identified during their research. In their paper, they present four categories, as can be seen in table II. The focus of this list is slightly different. While Chung et al. [6] analysed the whole Alexa environment, Li et al. [12] focused on data associated with a specific Amazon Echo device. Therefore, their findings present a subset of the data identified by Chung et al.

| Category | Description | Examples |
|---|---|---|
| Account | Information about the currently logged in Amazon account | name, email address, user ID |
| Alexa-enabled Device | A list of all registered devices that use the Alexa Voice Service (Echo devices), including basic device information | name, serial number, software version |
| Customer Setting | Device-independent settings associated with the current user account | Saved Wifi credentials, third-party services, location data |
| Skill | All the skills, both custom-made and from the Alexa skill store that are connected to the account | N/A |
| Compatible Device | List of devices that are connected to Alexa | UUID, name, network state |
| User Activity | All interaction between the user and the connected devices | Conversations, to-do lists, music playlists, notifications |
| Audio Data | Voice recordings of the user | N/A |

Table I
DATA CATEGORIES DESCRIBED BY CHUNG ET AL. [6]

| Category | Description | Examples |
|---|---|---|
| Device Data | All information about the device as such | Name, serial number, timezone, region |
| Connectivity | Information about the device's network connections | Wifi credentials, Bluetooth pairings, IP and MAC addresses |
| User Data | Data related to the functionality of the device | Calendars, lists, alarms, history, music |
| Application Data | Data regarding the software running on the Echo | user credentials, versions, Client ID, Product ID |

Table II
DATA CATEGORIES DESCRIBED BY LI ET AL. [12]

## III. METHODOLOGY

The focus of this paper is collecting and interpreting evidence specifically coming from the various APIs that the Alexa voice service provides. The first goal of this research was to determine specifically which data is stored on Amazon's servers when using the Alexa voice service on an Amazon Echo device. To do this, an example Echo device was populated with simulated user data. The seven categories of data established by Chung et al. [6] were used as a basis for this, which helped to make sure that all areas of the device's functionality are covered. The data was injected using specific crafted voice commands and menu interactions to make sure that the results were reproducible and to keep it as realistic as possible. After executing these actions and therefore filling the device with data, all known Alexa APIs were accessed to retrieve any data stored by Amazon, and the API responses were saved locally in the format they were delivered in.

Extracting the dataset from the Alexa API turned out to be a very complex and time-consuming task. To be able to do this, the Burp Suite software was used to intercept data packets being sent from Amazon's servers to the local browser when accessing the Alexa Web page. An obstacle for this was that Amazon encrypts all communication with this Web page with TLS. Additionally, it uses HTTP Strict Transport Security (HSTS), which prevents downgrade attacks [13]. This made it necessary to install the Root CA certificate that Portswigger [14] provides for this situation, as it enables Burp Suite to generate trusted server certificates when intercepting TLS connections. Burp Suite and Google Chrome were then set up in such a way that all communication with the domains alexa.amazon.com and alexa.amazon.co.uk was intercepted by Burp Suite and stored. The general infrastructure of this proxy connection can be seen in Figure 1. Then, the browser was used to manually access every available menu and functionality on the Web application, which subsequently caused it to request data from every relevant API. This was essentially active Web scraping of the Alexa Website. The proxy connection was used to automatically store all the raw JSON responses of any API that was accessed through the Website. The main points of interest here were the URL of the request and the response payload. Since all Alexa API calls return data in JSON format, it was straightforward to filter out the relevant packets in Burp Suite through their MIME type. All these packets were then exported to a file. Burp Suite's export function generates an XML file structure that contains the requests and responses including all metadata and payloads from the selected packets, which meant that a lot of the information was irrelevant for this purpose. Also, the file structure was hard to use, since it was comprised of single JSON structures for all the responses embedded in XML. Using a Python script, this data was converted into a simpler JSON format. The resulting file was then manually cleaned up. Duplicates were removed, as well as APIs that either did not contain any relevant data in the response, or did not return any response. The raw data files that were produced in this manner were too large to be included directly in this paper. They can however be accessed on GitHub under https://github.com/cleminator/AlexaPaperDatasets. This method uncovered a list of APIs, some of which have not yet been mentioned in the literature presented in section II-B. Table III lists all the APIs found. However, only a subset of them will be analysed in more detail in section IV.

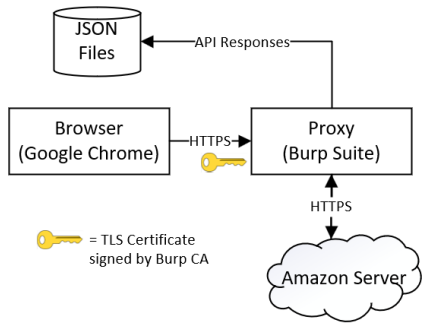

Figure 1. Proxy Infrastructure

| APIs | |
|---|---|
| /api/activities | /api/kedevice |
| /api/activity/privacy-link | /api/language |
| /api/allowed-providers | /api/lemur/access/ |
| /api/amazon-music-benefits | /api/lists/fetchUserPreference |
| /api/app-version-info | /api/lists/linkedPartners |
| /api/available-mid-field | /api/lists/listPartners |
| /api/bluetooth | /api/media/historical-queues |
| /api/bootstrap | /api/metrics-batch |
| /api/cards | /api/music-account-details |
| /api/communications/providers | /api/music/curated |
| /api/customer-status | /api/music/settings |
| /api/detect-first-run-devices | /api/namedLists |
| /api/device-preferences | /api/namedLists/ZZZ/items |
| /api/device-wifi-details | /api/notifications |
| /api/devices-v2/device | /api/np/player |
| /api/dnd/device-status-list | /api/np/queue |
| /api/dnd/schedule | /api/partner-authorization/details |
| /api/feature-alert | /api/phoenix |
| /api/feature-alert-location | /api/salmon/preferences |
| /api/featureaccess-v3 | /api/server-image |
| /api/gadgets/XXX/YYY/device-gadgets | /api/strings |
| /api/get-customer-pfm | /api/third-party |
| /api/get-languages | /api/traffic/settings |
| /api/video-skills/videoSkills | /api/wake-word |
| /api/household | /api/wake-words-locale |

Table III
DISCOVERED APIs

As mentioned before, Amazon Echo is a product family that consists of multiple different device types. It was not clear whether the choice of device made a difference for how data was processed. Therefore, these actions were performed with two different Amazon Echo devices independently. The same dataset was injected into both devices, so any differences in the respective API responses would indicate functional differences between them. In this case, two Amazon Echo Dots were used, one from the $2^{nd}$ and one from the most recent $3^{rd}$ generation. Separate Amazon Alexa accounts were created for each device, even though it would have been possible to connect multiple devices to the same account. This was done because an Echo device might also change device-independent settings, which could lead to scenarios where one device could overwrite data from the other. There are several differences between the two models. However, according to Digital Trends [15], these changes are all in hardware. While these changes may be interesting from a consumer's perspective, they should not make any difference to how data is processed. Since the software appears to be the same on both devices, the expected result was that there were no differences in the respective API responses. Comparing the two datasets mostly confirmed this. Table IV shows the an overview over the differences. While certain differences could be found, they were quite subtle, and could be attributed to slight differences in usage behaviour while populating the two devices with data as well as different timestamps, IDs, etc. Additionally of course, certain numbers and identifiers were different between the two, which can be attributed to the fact that they are different devices. In addition to the brief overview in table IV, the full data sets obtained from both devices can be found on GitHub under https://github.com/cleminator/AlexaPaperDatasets. Due to their size it was not possible to include them directly.

| API / Information | 2nd Gen | 3rd Gen |
|---|---|---|
| /device-preferences | - | Contains locale it-IT |
| /media/historical-queue | - | Different order of music items |
| /media/provider-contenttype-capabilities | - | Different order of music providers |
| /namedLists | - | Different IDs |
| /notifications | - | Different IDs |
| /phoenix/group | ApplianceGroup "Bedroom" is defined | - |
| /activities | - | Different items & order |
| /cards | - | Different items & order |
| Software Version | 641574820 | 2584225924 |
| Device Type | A3S5BH2HU6VAYF | A32DOYMUN6DTXA |
| Device Serial No. | G090*****0W3T | G090*****02GD |
| Device Account ID | A098**********G4IM | A072**********1ZU0 |
| Customer ID | ALU****B42 | A1P*****LPH |
| Customer Email | di*******4@gmx.net | di*******2@gmx.net |

Table IV
ECHO VERSION DIFFERENCES

This result shows, that while there are some differences between device generations, they do not seem to be relevant in the context of this research. As such, the remainder of this paper focuses on the most recent device, the 3rd generation. Finally, it can be assumed that this research will stay relevant in the future, as it seems that changes in hardware for Amazon Echo devices do not significantly impact data processing and storage. Because of this conclusion, the decision was made to focus only on the data extracted from the $3^{rd}$ generation Echo device from that point on.

## IV. API Data Analysis

Based on the API responses gathered through the method described above, this section contains a list of Alexa APIs together with their functionality. Additionally, possible artefacts and conclusions to be drawn from them are presented. This section is supposed to give an overview of certain artefacts to be found in Alexa API responses. In practice, this will be highly dependent on the specific situation, so certain aspects described here may not apply. This information can, however, be used as a baseline of possible areas to investigate.

### A. Activities

*API: https://alexa.amazon.com/api/activities*
This API returns a list of previous voice interactions with Amazon Alexa on the account that is currently logged in. Each entry contains several pieces of information, including a transcript of the command, a timestamp, information about the device the command was given to and data about the given response. Additionally, a so-called utterance ID is included, which is linked to the actual audio recording of the command.

Since this API summarises all communication between the user and Alexa, it can provide an investigator with a good insight into the user's behaviour, habits and possibly also whereabouts. Certain commands may contain location information that can be useful in an investigation. When a user asks for information about the weather, for example, Alexa responds with data about a specific location. If the user does not specify a location, Alexa uses the one that is set as default, which is usually where the device is placed. This location information combined with the time stamp included can be a hint towards a person's whereabouts at a specific time. While the user could potentially issue commands through his smartphone from anywhere, the investigator can find out specifically from which device the command was given through the device ID that is part of the response. Other commands, such as asking for traffic information may provide similar conclusions. This API is essential for understanding the user's behaviour around his Echo devices. Because the response is likely to be very large and hard to read depending on the number of interactions, it may be sensible to visualise the events as a timeline.

As part of each entry, a value called "activityStatus" is provided. For successful interactions, this value is set to "SUCCESS". In some cases, however, entries may have the status "DISCARDED_NON_DEVICE_DIRECTED_INTENT". This means that an Echo device started recording because it detected the defined wake word, which then turned out to be a false positive. Even though in this case no command was executed, there are still voice recordings associated with it, which may contain sensitive information through background noise, as well as indicate the suspect's whereabouts.

### B. Bluetooth

*API: https://alexa.amazon.com/api/bluetooth*
Amazon provides the possibility to connect Echo devices to external speakers through Bluetooth. The information about paired devices can be accessed through this API. It returns a list of devices including their respective pairings and contains details such as the device's serial number, device type, and friendly name, as well as the friendly names of the Bluetooth devices and their current connection states.

In certain instances, the information about the connected devices may be of interest, as it can potentially be used to identify further devices to be analysed.

### C. Bootstrap & Household

*APIs:*
*https://alexa.amazon.com/api/bootstrap*
*https://alexa.amazon.com/api/household*
These two APIs return basic information about the account that is logged in. Bootstrap simply contains the registered name of the user, their email address and a customer ID. More than one user can be associated with an account. Household provides some more details for each of them, namely their first and full names, a parameter called "role", which can be "ADULT" or "CHILD" and their respective IDs.

These IDs are included in several other APIs, such as Activities, and allow an investigator to associate interactions with a specific user.

### D. Cards

*API: https://alexa.amazon.com/api/cards*

Recent interactions with Alexa are displayed in a tile-based layout of cards by both the Web and mobile management applications, with each card corresponding to a particular action. There is a lot of overlap between this API and activities, as they both contain the same data, but formatted differently.

### E. Contacts

*API: https://alexa-comms-mobile-service.amazon.co.uk/user/XYZ/contacts*

This API returns a list of all contacts registered in the address book of the current Alexa account. This includes their name, phone numbers, email addresses, postal address and more.

There are two different types of contacts to be found here. The user can manually add contacts to their address book, or give the Alexa companion application permission to import all contacted from their smartphone. Permissions are acquired at login time. This can be very valuable information for an investigator, as it might give him access to a suspect's contacts without needing to access their phone.

### F. Device Preferences

*API: https://alexa.amazon.com/api/device-preferences*

As the name suggests, this API returns information about all Echo devices registered to the account. First of all, there is data about the device itself, such as its serial number, device type, and the associated account ID. Secondly, detailed information about the device's location and language settings is included in the form of a locale and time zone setting, and an address field comprised of country, county, city, postal code, street, and number. Additionally, it contains settings for temperature and distance units.

While the information might be useful to an investigator, it should be noted that the user can manually adjust these settings, so they should not be blindly trusted. Instead, they should be cross-referenced with location information from other APIs, such as Activities.

### G. WiFi Details

*API: https://alexa.amazon.com/api/device-wifi-details*

When invoked with a device's serial number as a parameter, this API responds with information about the wireless network settings of said device. Specifically, the response parameters are the device's MAC address, its serial number and device type, as well the ESSID of the network it is connected to.

Even though this is not very detailed, there are still conclusions to be drawn from this information. The information on which wireless network the device is connected to can be used to gather further information. By knowing which network was used, the investigator can attempt to collect information from other devices in that network, or possibly even gather network logs that might contain more evidence. The fact that the Echo device's MAC address is also included here can help to identify traffic coming specifically from this device. It should be noted, however, that the traffic between an Echo device and Amazon's servers is usually encrypted.

### H. Devices

*API: https://alexa.amazon.com/api/devices-v2*

Similarly to the Device Preferences API described above, this one provides details about the devices registered to the account. Compared to Device Preferences, this API contains more technical data, which may be interesting. The parameters include the device's MAC address, device family, software version, device type, whether it is online at the moment, its friendly name and whether it is currently being charged.

The information about the device's type and software version could be used by an investigator to identify known weaknesses or vulnerabilities that may allow them to directly access the device and gather further evidence. In some cases, several devices might be registered to a single account, in which case the information from this API can support an investigator's efforts to identify them and to associate their preferences with the physical devices. This could be done for example through the MAC address or device type.

### I. Named Lists

*API: https://alexa.amazon.com/api/namedlists*

As the name already suggests, this API can be used to gather data about any lists that are kept in the account. Two default lists always exist, namely "To-do" and "Shopping". However, other lists with custom names can be arbitrarily created by the user. Each list contains several items, which can be marked "completed". Both the lists and the entries in them are identified by an ID. The API takes a list ID as a parameter and returns only data about that list. The API response contains all entries associated with the list, which includes their respective IDs, names, whether they are marked completed, as well as timestamps for when they were created and last updated. If no parameter is given, the API responds with the IDs and names of all lists, as well as creation and update timestamps for the whole lists, rather than single items.

Since these lists can be used for any purpose, their usefulness to an investigator highly depends on how much and for what the suspect was using them. Generally, the timestamps may be of interest, as they allow the entries to be included in a general event timeline, where they might provide more context to the situation.

### J. Phoenix

*API: https://alexa.amazon.com/api/phoenix*

The Amazon Alexa ecosystem allows users to place several devices in their homes to form a complete smart home system. For management purposes, Amazon provides the functionality to create different groups of devices, assign devices to certain rooms, associate Echo devices with connected appliances and more. The Phoenix API returns this structure.

An investigator could use that information to find further devices that might be relevant to the investigation. For example, if an investigator is trying to establish what happened in a certain room in a suspect's house, they could use this API's response to find other devices in that same room.

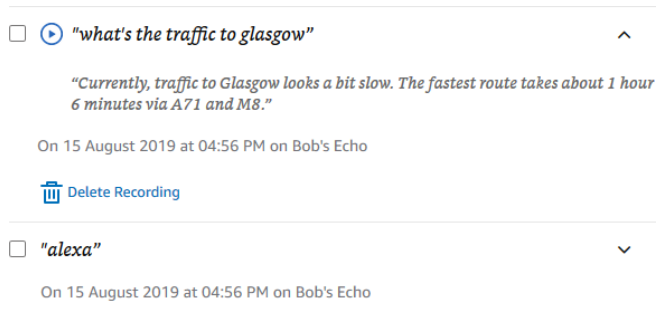

Figure 2. Alexa Voice History

### K. Utterance

*API: https://alexa.amazon.com/api/utterance/audio/data*

This API is different from the others because it is the only one that does not return data in JSON format. When a user gives a voice command to an Echo device, that command is recorded and sent to Amazon's server for processing. There, the intent behind the command is detected, a response formulated and returned to the device. The voice recordings are then stored on the servers indefinitely, as recently confirmed by Amazon [16]. The utterance API can be used to access any voice recording associated with the currently authenticated account. The audio files are referenced through an utterance ID, which can be found in the entries of the Activities API, as described above.

These recordings can be of great interest to an investigator because they allow them to tie interactions with the device to a physical person through their voice. If a device has multiple users, the audio can be used to discern who triggered which commands. Furthermore, in certain cases, the recordings might contain additional information in the form of background noise. Allegedly, Amazon employees found a recording where sexual assault could be heard in the background, which shows the potential relevance of this [1].

## V. User Privacy Aspects

Amazon provides a number of options for users to control certain privacy aspects of their accounts. As the last part of this research, these settings were reviewed.

In the Amazon Alexa settings menu, there is a specific section for *Alexa Privacy*[1]. In this section, the user can manage certain privacy aspects around Alexa and their Echo devices, as well as access the *Amazon Privacy Note* which explains what Amazon may do with personal information. The first option, *Review Voice History*, allows the user to see a list of all recent voice interactions with Alexa. The user can choose a date range to filter entries. For each entry, a transcription of the voice command is shown along with a timestamp and the device it was given to. The user can also listen to the audio recordings associated with the commands. Figure 2 shows an excerpt of this menu. Additionally, it is possible to delete entries from the interaction history. This can be done for single entries, certain time frames or the whole history.

To test whether deleting entries from the Alexa interaction history actually removes the data associated with them, several scenarios were tested. For each case, a new voice command such as *"What's the weather in Edinburgh?"* was given to the Echo device. After that, the respective entry was accessed through the *Activities* API, the *Cards* API, the *Voice History* menu in the Alexa privacy settings and the *Utterance* API.

For each scenario, the entry was deleted through one of the methods described below. Then, all APIs were accessed again to find out whether the data was actually deleted.

The settings of the Alexa Web page contain a menu called *History*[2], that shows all recent interactions and lets the user delete them (figure 3). Deleting an entry from this view made the biggest impact. After that, the entries in both the *Activities* and the *Cards* API were no longer available via the API. Additionally, the associated voice recording was also deleted.

Each card on the homepage of both the Alexa Web page and companion app has a *Remove card* button with which they can be deleted. When the respective card was deleted, that card was no longer returned by the API, nor shown on the Alexa Web page or app. The entry in the *Activities* API and the audio recording were still available and unchanged.

As described above, the Alexa privacy menu contains the option to delete voice recordings (figure 2). When the voice recording was deleted through this option it was no longer available via the *Utterance* API However, the entries in both the *Activities* and the *Cards* API were still available, including any metadata associated with it. Therefore, deleting entries from the voice history in the Alexa privacy settings specifically only removes audio recordings, and not any metadata.

None of the other known APIs were changed by any of these actions. Even if an entry is fully deleted, it may therefore still be possible to reconstruct the action that was performed using timestamps and other metadata from other APIs. If a suspect, for example, asked Alexa to add certain suspicious items to their shopping list using a voice command, and then deleted this interaction using the methods above to cover their tracks, an investigator could still use the information returned by /api/namedlists to acquire these items as well as the timestamps from when they were added and last modified. Using the fact that certain information is stored redundantly and available through multiple APIs can be a legitimate strategy against anti-forensics attempts.

## VI. Conclusions

The first goal of this research was to determine how Amazon processed and stored data about the users of Amazon Echo devices. To do this, the test device was first populated with simulated user data by issuing certain voice commands to the device and changing settings in the Alexa companion application. To determine the format in which that data was stored, it was then retrieved again through a list of APIs. By crawling through the Alexa Web application and recording any calls to an API using Burpsuite, additional APIs could be found. All responses from the known APIs were then downloaded and correlated with the dataset that was injected into the device in the first place. To determine whether the choice of device hardware somehow influenced the structure of the data, this was done using the Amazon Echo Dot's 2$^{nd}$ generation and 3$^{rd}$ generation. When comparing the two results it was determined that there were no significant changes in the datasets. This lead to the conclusion that the choice of hardware did not make a difference to the data stored by Amazon. This also indicated that the results of this research will stay relevant in the future including future releases of the device. Because there is no official documentation of the APIs, all knowledge about them comes from exploratory work. In the future, Amazon might change their structure to accommodate new features, which means that this type of research must be performed regularly for investigators to keep being able to use them. However, it was shown that new hardware versions to not affect the APIs, which allows investigators to perform the same type of analysis for multiple devices. The evaluation of the APIs also showed that a lot of information was stored in multiple places, spread over different APIs. An investigator can potentially use this to determine whether a suspect tried to delete data to cover their tracks. Generally, most of the APIs store metadata such as timestamps, device information or location data. This facilitates the investigators' ability to determine an event timeline. Section IV provides investigators with a list of known APIs, as well as a detailed analysis of potential artefacts that can be found in their respective responses. Some of these will not be relevant for certain scenarios, but they generally provide the investigator with valuable information about how to gather digital evidence.

---

[1] https://www.amazon.co.uk/hz/mycd/myx\#/home/alexaPrivacy/home

[2] https://alexa.amazon.co.uk/spa/index.html\#settings/dialogs

Figure 3. Alexa Interaction History

Finally, Amazon's privacy settings were reviewed. Out of the options presented to the user, the *Voice History* is the most interesting one. It allows the user to review all past voice commands issued to Alexa, including the voice recordings. The user can also delete selected entries or whole ranges. The functionality presented by this menu is similar to the two APIs *Cards* and *Activities*, both of which also show interaction history in different formats, as described in section IV. It is possible for the user to delete entries from any of these lists, which led to the question of whether there are any differences in the data that is actually deleted. It was shown that deleting an entry through the *History* menu in the settings of the Alexa Web pageled to the most comprehensive deletion. However, it should be noted that entries in other APIs are unaffected by this, which might still enable someone with access to all APIs to make conclusions about the user's behaviour. This is a very important insight for investigators because it means that even if a suspect tries to cover their tracks, some artefacts may still be available. Also, because a lot of the information is somewhat redundant as it is available through more than one API, the investigator might even be able to detect attempts of anti-forensics techniques.

REFERENCES

[1] M. Day, G. Turner, and N. Drozdiak, "Amazon Workers Are Listening to What You Tell Alexa," 2019. [Online]. Available: https://www.bloomberg.com/news/articles/2019-04-10/is-anyone-listening-to-you-on-alexa-a-global-team-reviews-audio

[2] K. Paul, "Google workers can listen to what people say to its AI home devices," 2019. [Online]. Available: https://www.theguardian.com/technology/2019/jul/11/google-home-assistant-listen-recordings-users-privacy

[3] Gartner, "Gartner Says Worldwide Spending on VPA-Enabled Wireless Speakers Will Top $3.5 Billion by 2021," 2017. [Online]. Available: https://www.gartner.com/en/newsroom/press-releases/2017-08-24-gartner-says-worldwide-spending-on-vpa-enabled-wireless-speakers-will-top-3-billion-by-2021

[4] A. Cuthbertson, "Amazon Ordered to Give Alexa Evidence in Double Murder Case," 2018. [Online]. Available: https://www.independent.co.uk/life-style/gadgets-and-tech/news/amazon-echo-alexa-evidence-murder-case-a8633551.html

[5] M. Sampathkumar, "Amazon Echo could become key witness in murder investigation after data turned over to police," 2017. [Online]. Available: https://www.independent.co.uk/news/world/americas/amazon-echo-murderinvestigation-data-police-a7621261.html

[6] H. Chung, J. Park, and S. Lee, "Digital forensic approaches for Amazon Alexa ecosystem," *Digital Investigation*, vol. 22, pp. S15–S25, 2017.

[7] A. C. Service, "Alexa, Echo Devices, and Your Privacy." [Online]. Available: https://www.amazon.co.uk/gp/help/customer/display.html/?nodeId=GA7E98TJFEJLYSFR

[8] Digital Forensic Research Workshop, "A Road Map for Digital Forensic Research," in *The Digital Forensic Research Conference DFRWS 2001 USA*, Utica, New York, 2001.

[9] K. Kent, S. Chevalier, T. Grance, and H. Dang, "Special Publication 800-86 Guide to Integrating Forensic Techniques into Incident Response," NIST, Tech. Rep., 2006.

[10] D. A. Orr and L. Sanchez, "Alexa, did you get that? Determining the evidentiary value of data stored by the Amazon® Echo," *Digital Investigation*, vol. 24, pp. 72–78, 2018.

[11] O. Piette, "The Amazon Echo API," 2014. [Online]. Available: https://www.piettes.com/the-amazon-echo-api/

[12] S. Li, K.-k. R. Choo, Q. Sun, and W. J. Buchanan, "IoT Forensics: Amazon Echo as a Use Case," *IEEE Internet of Things Journal*, vol. 14, no. 8, pp. 1–11, 2015.

[13] J. Hodges, C. Jackson, and A. Barth, "RFC 6797 - HTTP Strict Transport Security (HSTS)," Internet Engineering Task Force, Tech. Rep., 2012.

[14] Portswigger, "Installing Burp's CA Certificate in your browser." [Online]. Available: https://support.portswigger.net/customer/en/portal/articles/1783075-Installing_Installing%20CA%20Certificate.html

[15] E. Rawes, "Second-gen vs. third-gen Echo Dot: What's the difference?" 8 2019. [Online]. Available: https://www.digitaltrends.com/home/amazon-echo-dot-gen-2-vs-gen-3/

[16] O. Tambini, "Amazon Alexa stores voice recordings for as long as it likes (and shares them too)," 2019. [Online]. Available: https://www.techradar.com/news/amazon-alexa-stores-voice-recordings-for-as-long-as-it-likes-and-shares-them-too